\documentclass[%
 reprint,
superscriptaddress,
 amsmath,amssymb,
 aps,
pra,
]{revtex4-1}

\usepackage{latexsym,amsmath,amssymb}
\usepackage{braket} 
\usepackage{ftnxtra}
\usepackage{fnpos}
\usepackage{tikz}
\usepackage{color}
\usepackage{footnote}
\newcommand{\xinhui}[1]{\textcolor{black}{#1}}

\newcommand{\nn}{\nonumber}

\usepackage{amsthm} 
\usepackage{appendix}

\newcommand{\beq}{\begin{equation}}
	\newcommand{\eeq}{\end{equation}}
\newcommand{\bqa}{\begin{eqnarray}}
	\newcommand{\eqa}{\end{eqnarray}}

\newcommand{\rt}[1]{\sqrt{#1}\,}

\newcommand{\sch}{Schr\"odinger}

\newcommand{\sq}[1]{\left[ {#1} \right]}

\newcommand{\an}[1]{\left\langle{#1}\right\rangle}
\newcommand{\tr}[1]{{\rm Tr}\sq{ {#1} }}

\usepackage[T1]{fontenc} 

\usepackage{hyperref}
\hypersetup
{
colorlinks=true,
linkcolor=blue,
filecolor=blue,
urlcolor=blue,
citecolor=cyan,
}

\usepackage{graphicx}

\newcommand{\ba}{\begin{eqnarray}}
\newcommand{\ea}{\end{eqnarray}}
\newcommand{\ban}{\begin{eqnarray*}}
\newcommand{\ean}{\end{eqnarray*}}

\begin{document}

\title{Device-independent verification of Einstein-Podolsky-Rosen steering }
\author{Yuan-Yuan Zhao}
\thanks{These authors contribute to this work equally.}
\affiliation{College of the Science, China University of
Petroleum, 266580 Qingdao, China\\
$^2$ National Laboratory of Solid State Microstructures, School of Physics,
and Collaborative Innovation Center of Advanced Microstructure, Nanjing University, Nanjing, 210093, China}
\affiliation{Peng Cheng Laboratory, Shenzhen 518055, China}

\author{Chao Zhang}
\thanks{These authors contribute to this work equally.}
\affiliation{College of the Science, China University of
Petroleum, 266580 Qingdao, China\\
$^2$ National Laboratory of Solid State Microstructures, School of Physics,
and Collaborative Innovation Center of Advanced Microstructure, Nanjing University, Nanjing, 210093, China}
\affiliation{Synergetic Innovation Center of Quantum Information and Quantum Physics, University of Science and Technology of China, Hefei, 230026, China}

\author{Shuming Cheng}
\email{shuming\_cheng@tongji.edu.cn}
\affiliation{Department of Control Science and Engineering, Tongji University, Shanghai, 201804, China}
\affiliation{Institute for Advanced Study, Tongji University, Shanghai, 200092, China}
\affiliation{Institute for Quantum Computing, Baidu Research, Beijing, 100193, China}

\author{Xinhui Li}
\email{lixinhui@nju.edu.cn}
\affiliation{State Key Laboratory of Networking and Switching Technology,Beijing University of Posts and Telecommunications, Beijing, 100876, China}
\affiliation{National Laboratory of Solid State Microstructures and School of Physics, Nanjing University, Nanjing 210093, China}
\author{Yu Guo}
\affiliation{College of the Science, China University of
Petroleum, 266580 Qingdao, China\\
$^2$ National Laboratory of Solid State Microstructures, School of Physics,
and Collaborative Innovation Center of Advanced Microstructure, Nanjing University, Nanjing, 210093, China}
\affiliation{Synergetic Innovation Center of Quantum Information and Quantum Physics, University of Science and Technology of China, Hefei, 230026, China}
\author{Bi-Heng Liu}
\email{ bhliu@ustc.edu.cn}

\affiliation{College of the Science, China University of
Petroleum, 266580 Qingdao, China\\
$^2$ National Laboratory of Solid State Microstructures, School of Physics,
and Collaborative Innovation Center of Advanced Microstructure, Nanjing University, Nanjing, 210093, China}
\affiliation{Synergetic Innovation Center of Quantum Information and Quantum Physics, University of Science and Technology of China, Hefei, 230026, China}
\author{Huan-Yu Ku}
\affiliation{\xinhui{Faculty of  Physics, University of Vienna,  Boltzmanngasse 5, 1090 Vienna, Austria }}
\affiliation{\xinhui{Institute  for  Quantum  Optics  and  Quantum  Information (IQOQI), Austrian  Academy  of  Sciences,  Boltzmanngasse 3, 1090 Vienna, Austria}}
\author{Shin-Liang Chen}
\affiliation{\xinhui{Department of Physics, National Chung Hsing University, Taichung 40227, Taiwan}}
\author{Qiaoyan Wen}
\affiliation{State Key Laboratory of Networking and Switching Technology,Beijing University of Posts and Telecommunications, Beijing, 100876, China}
\author{Yun-Feng Huang}

\author{Guo-Yong Xiang}
\email{ gyxiang@ustc.edu.cn}
\author{Chuan-Feng Li}
\email{ cfli@ustc.edu.cn}
\author{Guang-Can Guo}
\affiliation{College of the Science, China University of
Petroleum, 266580 Qingdao, China\\
$^2$ National Laboratory of Solid State Microstructures, School of Physics,
and Collaborative Innovation Center of Advanced Microstructure, Nanjing University, Nanjing, 210093, China}
\affiliation{Synergetic Innovation Center of Quantum Information and Quantum Physics, University of Science and Technology of China, Hefei, 230026, China}

\begin{abstract}
 Entanglement lies at the heart of quantum mechanics, and has been identified an essential resource for diverse applications in quantum information. If entanglement could be verified without any trust in the devices of observers, i.e., in a device-independent (DI) way, then the high security can be guaranteed for various quantum information processing tasks. In this work, we propose and experimentally demonstrate a DI protocol to certify the presence of entanglement based on Einstein-Podolsky-Rosen (EPR) steering. We first establish the DI verification framework by taking the advantages of measurement-device-independent technique and self-testing, which is able to verify all bipartite EPR-steerable states. In the scenario of three-measurement settings for per party, the protocol is robust in tolerance of inefficient measurements and imperfect self-testing. Moreover, a four-photon experiment is implemented for verification beyond  Bell nonlocal states. Our work presents the new insight into quantum physics and  paves the way for realistic implementations of  secure quantum information  processing tasks.  
\end{abstract}


\maketitle
\smallskip

\section{Introduction}
Entanglement is of fundamental importance to understand quantum theory, and also has found wide applications in quantum communication and computation tasks~\cite{Horodecki2009}. If its presence could be certified without imposing any trust in the involved parties and their devices, then it is likely to guarantee information processing tasks with unconditional security. As a celebrated example, Bell inequalities~\cite{Bell1964,Brunner2014} violation offers such a device-independent (DI) protocol. However, the conclusive violation of Bell inequalities typically requires a high efficiency of measurement apparatuses to close the detection loophole. Besides, it also demands the low transmission loss since sufficiently lossy entangled states are unable to violate any Bell inequality~\cite{Werner1989}. Thus, the practical utility of this DI verification based on Bell inequalities is compromised in noisy quantum networks.

Notably, Bowles {\it et al.} has shown in~\cite{Bowles2018a,Bowles2018b} that combining the measurement-device-independent (MDI) technique~\cite{Buscemi2012} with self-testing~\cite{Mayers2004,Supic2019,Chen2021} yields an alternate way which is able to device-independently verify all entangled states and circumvent the potential detection loophole~\cite{Branciard2013}. However, its complete implementation relies on the near-perfect self-testing of a set of prepared states with average fidelity above $99.998\%$~\cite{Bowles2018b}, making it unrealistic to implement within current technology.

In this work, we propose an experimental-friendly DI  protocol free of all above limitations (See Fig.~\ref{fig:DISM}), based on Einstein-Podolsky-Rosen (EPR) steering~\cite{Wiseman2007,Jones2007,Xiang2022}. Quantum steering was introduced by \sch ~to describe the ability that if certain pure entangled state is shared by two observers, one can remotely prepare the other's states via choosing suitable measurements~\cite{schrodinger1935,Quintino2014,Ku2022,2022Ku}. It was operationally reformulated as EPR-steering via a task of verifying entanglement by Wiseman {\it et al.}~\cite{Wiseman2007}. Since it lies strictly intermediate between entanglement and Bell nonlocality~\cite{Wiseman2007,Quintino2015}, this hierarchy implies that EPR-steering is experimentally less demanding than Bell nonlocality to verify entanglement, confirmed in various experimental setups~\cite{Saunders2010,Bennet2012,Smith2012,Handchen2012,Peise2015,Kunkel2018,Wang2018,Fadel2018}. Moreover, the possibility of MDI verification of steering have been shown~\cite{Cavalcanti2013,Kocsis2015,Guo2018,Jeon2019,Zhao2019,Bartosik2021}, together with experimental validations reported in~\cite{Kocsis2015,Guo2018,Zhao2019}. Hence, following from works~\cite{Bowles2018a,Bowles2018b}, using self-testing, we can establish a DI protocol to verify EPR-steering and hence entanglement.

We first show that all EPR-steerable states can be verified within this DI framework. Particularly, if three-measurement settings as per party are assumed, we obtain a steering inequality suitable for DI certification under imperfect self-testing with average fidelity \xinhui{$99.7\%$}, which is a significant reduction in comparison to the DI verification based on entanglement. Finally, we implement a proof of principle experiment with four photons to validate the DI steering protocol, and find it can even verify Bell local states with an experimentally attainable self-testing fidelity of \xinhui{ around $99.95\%$}.

\section{Preliminaries}
Suppose that two space-like separated observers, Alice and Bob say, make measurements on a preshared state. Denote Alice's and Bob's measurements $x$ and $y$ respectively, and the corresponding outcomes $a$ and $b$.  EPR-steering from Alice to Bob is demonstrated if the measurement statistics $p(a,b|x,y)$ cannot be explained by any {\it local hidden state} model as $p(a, b|x, y)=\sum_\lambda p(\lambda) p(a|x, \lambda) {\rm Tr}[E^B_{b|y}\rho^B_\lambda],$
where the hidden variable $\lambda$ specifies some classical probability distribution $p(a|x, \lambda)$ for Alice and some quantum probability distribution ${\rm Tr}[E^B_{b|y}\rho^B_\lambda]$ for Bob which is generated via performing a positive-operator-valued measurement $\{E^B_{b|y}\}_{b, y}$ on quantum states $\rho^B_\lambda$~\cite{Wiseman2007}. Note that Alice's side may not obey quantum rules, so EPR-steering is intrinsically an one-sided device-independent verification task. For any steerable state, the detection task can be accomplished via violating a linear steering witness of the form~\cite{Cavalcanti2009}
\begin{equation}
W_{\rm S}=\sum_{j}\an{a_jB_j}\leq 0. \label{Steerineq}
\end{equation}
Here $a_j$ corresponds to the outcome of Alice's measurement $j$, and $B_j$ represents Bob's $j$-th observable. 

\begin{figure}[htbp]
	\includegraphics[width=0.99\linewidth]{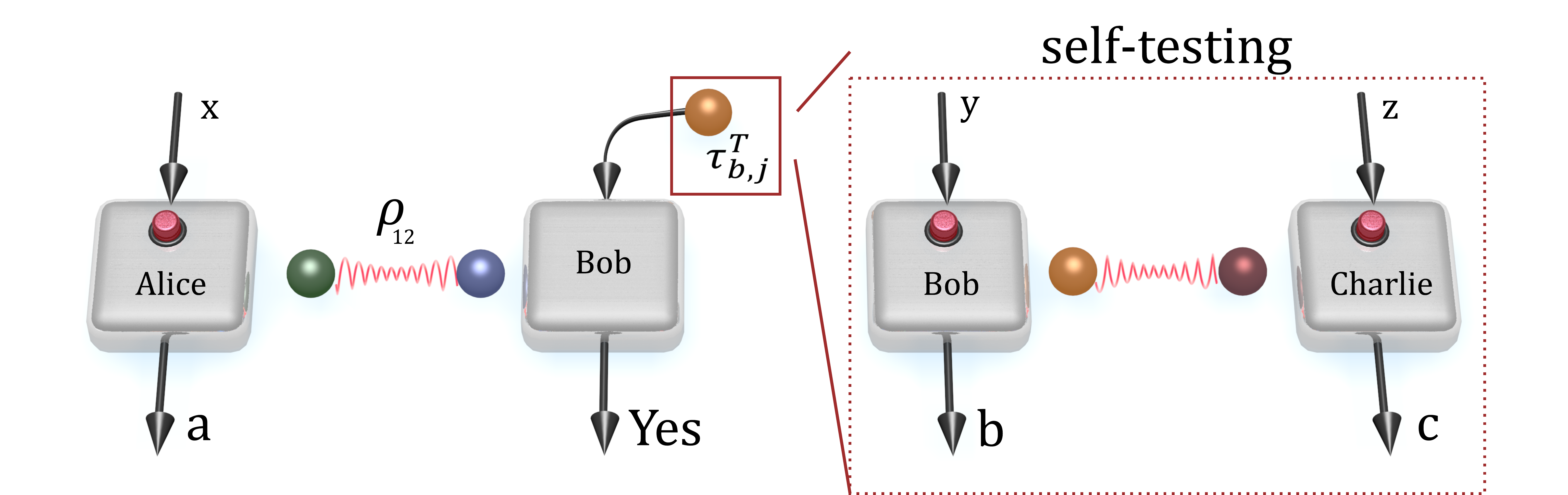}
	\caption{{\bf DI verification framework of EPR-steering.} The DI protocol is composed of two procedures. One is illustrated in the left side which corresponds to the MDI verification of the state $\rho_{12}$. In this step, Alice randomly takes measurements $x$ and obtains $a$, while Bob performs one binary measurement on his local system and a set of states $\{\tau_{b, j}^T\}$ assigned from Charlie, and collects the outcome ``{\rm Yes}".  The second is described in the right box, corresponding to the self-testing process. Noting $\tau^T_{b,j}$ can be prepared by Charlie performing local measurements $z_j=\{\tau_{b,j}\}$ on Bell state $\ket{\Phi^+_d}$ preshared by Bob and Charlie, this measurement strategy can be self-tested via the violation of Bell inequalities, such as the Bell-CHSH one used in the main text.  }\label{fig:DISM}
	\rule[-10pt]{8.5cm}{0.05em}
\end{figure}

Certifying the presence of EPR-steering can be adapted to the MDI scenario~\cite{Cavalcanti2013,Kocsis2015,Guo2018,Jeon2019} where the trust in Bob's devices required in Eq.~(\ref{Steerineq}) is transferred to a third observer, Charlie say, who prepares a set of quantum states and sends them at random to Bob. As in Fig.~\ref{fig:DISM}, upon receiving these states described by density matrices $\{\tau^T_{b,j}\}$  with $T$ being the transpose operation, Bob is required to perform an arbitrary binary measurement $\mathcal{B}$ with which the outcomes are modelled as either ``{\rm Yes}" or ``{\rm No}". Denote by $P(a,{\rm Yes}\,|x, \mathcal{B}, \tau^T_{b,j})$ the probability that Alice obtains $a$ for the measurement $x$ and Bob answers ``Yes'' when assigned to  $\tau^T_{b,j}$. Then, arranging the corresponding outcome statistics as Eq.~(\ref{Steerineq}) leads to a MDI steering witness~\cite{Kocsis2015,Guo2018}
\begin{equation}
W_{\rm MDI}=\sum_{a,b,j} g_{b,j} a_j P(a, {\rm Yes}\,|x=j ,\mathcal{B}, \tau^T_{b,j})\leq 0, \label{QRS}
\end{equation}
with $g_{b, j}$ being some predetermined weights. \xinhui{Typically, these coefficients can be chosen as the weights of measurement elements for Bob's observable $B_j=\sum_b g_{b, j}E_{b|j}$}. As the measurement outcome ``{\rm Yes}'' is only recorded, Bob's side allows for extremely low measurement efficiency~\cite{Branciard2013}.

The optimal measurement strategy for Bob is to perform a partial Bell state measurement (BSM) $\mathcal{B}\equiv\{\mathcal{B}_1, \mathbb{I}-\mathcal{B}_1\}$ where $\mathcal{B}_1 = | \Phi^+_d\rangle\langle \Phi^+_d|$ with $| \Phi^+_d\rangle=\sum_j|jj\rangle/\rt{d}$ models the answer ``Yes'' and $d$ is the dimension of the Hilbert space of $\{\tau^T_{b,j}\}$ equal to that of Bob's local system. Indeed, given an arbitrary steerable state, its MDI witness~\eqref{QRS} can be constructed from the corresponding witness~(\ref{Steerineq}), implying all steerable states are detectable in the MDI manner~\cite{Kocsis2015,Guo2018}.

\begin{figure*}[ht]
	\centering
	\includegraphics[width=0.9\linewidth]{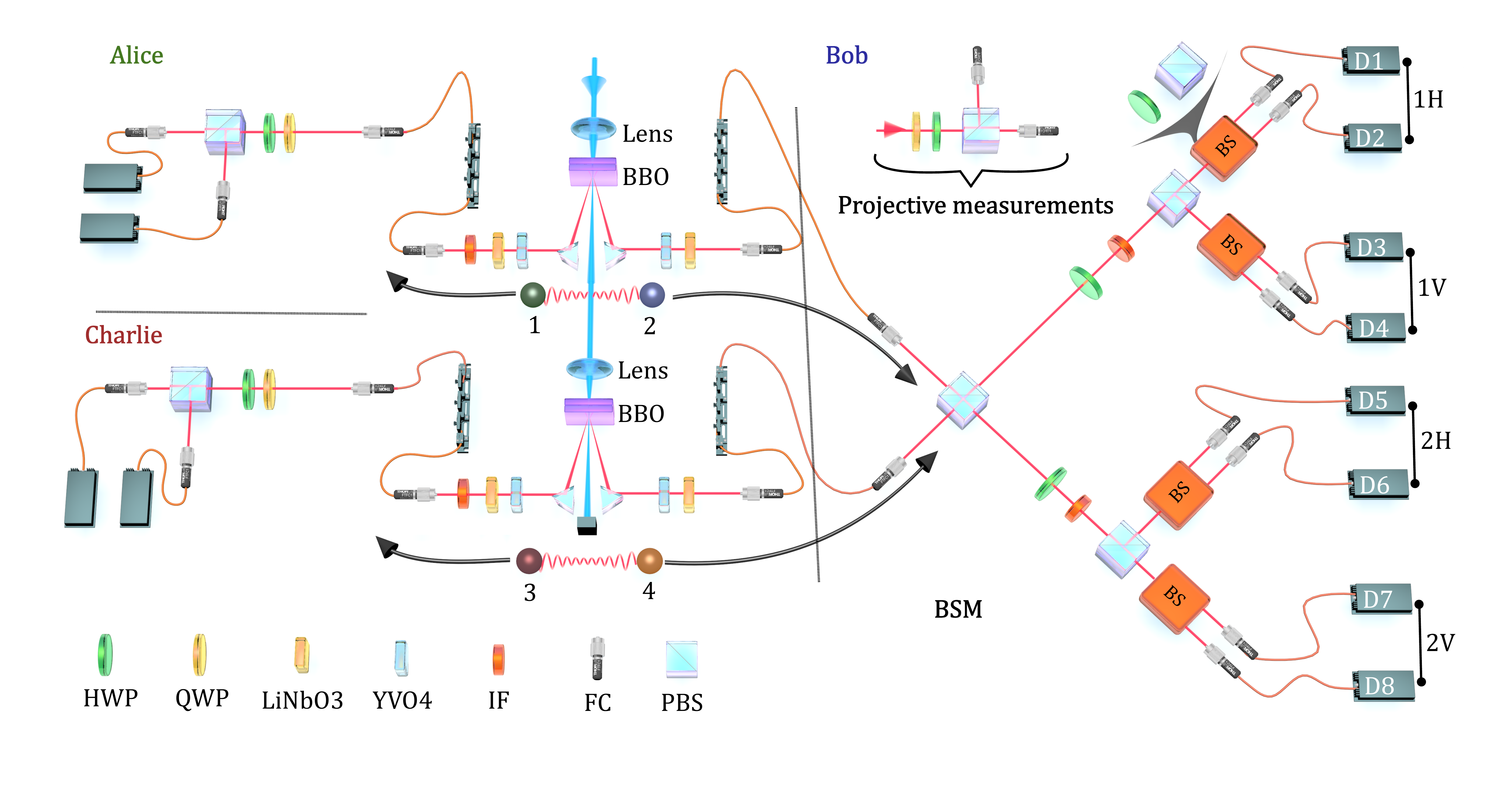}
	\caption{{\bf Experimental setup.} Two pairs of entangled photons are generated via the spontaneous parametric down-conversion process. A sandwich-like $\beta$-barium borate (BBO) crystal is configured to prepare entangled photons with high fidelity and high brightness. One pair labelled as $1$ and 2 distributed to Alice and Bob, is generated as Werner states~\eqref{Werner}, while the other labelled as $3$ and $4$ is produced as the Bell state $\ket{\Phi^+_2}$ sent to Charlie and Bob respectively. For the three-measurement case, a complete implementation of DI steering verification requires a triple Bell-CHSH test~\eqref{eq:CHSH} and the noisy DI steering test ~\eqref{noisy}. Alice and Charlie perform three Pauli measurements $\sigma_j$ on their respect photons, while Bob makes 6 measurements described by $(\sigma_i+\sigma_j)/\rt{2}$ and $(\sigma_i-\sigma_j)/\rt{2}$ on the photon 4 and an additional partial BSM $\mathcal{B}$ on his photons 2 and 4. Abbreviations of the components: HWP, half wave plate; QWP, quarter wave plate; \xinhui{LiNbO3, Lithium niobate crystal; YVO4, Yttrium vanadate crystal;} IF, interference filter; FC, fibre coupler; PBS, polarising beam splitter;  BSM, Bell state measurement; BS, beam splitter; D1-D8, single photon detector.}
	\label{fig:setup}
\rule[-10pt]{18cm}{0.05em}
\end{figure*}

\section{Device-independent verification of EPR steering} Within the MDI framework witnessed via Eq.~(\ref{QRS}), both Alice's and Bob's sides are already device-independent, except for Charlie who prepares quantum states for Bob. Consequently, eliminating trust in Charlie's devices yields a fully DI steering verification protocol. As discussed below, this can be accomplished via self-testing which aims to identify the states and measurements for completely uncharacterised devices~\cite{Mayers2004,Supic2019,Scarani2013}.

Note first that the states $\{\tau^T_{b,j}\}$ can be prepared by Charlie making local measurements $z_j=\{\tau_{b,j}\}$ on Bell state $\ket{\Phi^+_d}$ due to the relation ${\rm Tr}_C[\mathbb{I}^B\otimes \tau_{b,j}\ket{\Phi^+_d}\bra{\Phi^+_d}]= \tau^T_{b,j}/d$. This preparation process, including Bob's additional measurements $y$, can be uniquely determined or self-tested via the well-chosen Bell inequality of which its maximal violation is only achieved at each party performing a certain set of measurements on a specific state, up to some local isometry. Thus, using self-testing to determine the input states $\tau_{b, j}^T$ in Eq.~(\ref{QRS}), we can obtain a DI steering inequality as~\cite{SM}
\begin{equation}
W_{\rm DI}=\sum_{a,c,j} g_{c,j} a_j P(a, {\rm Yes}, c\,|x=j, \mathcal{B}, z=j)\leq 0. \label{DI0}
\end{equation}
Here Charlie making measurements $z$ and obtaining outcomes $c$ is equivalent to he sending a state $\tau_{b, j}^T$ to Bob, \xinhui{and $g_{c, j}$ are close relate to the weights $g_{b, j}$ in Eq.~\eqref{QRS}}.  We remark that the self-testing process, involving $\Ket{\Phi^+_d}$ and Charlie's measurements, is not explicitly assessed in the above DI witness~\eqref{DI0} and requires a detailed analysis case by case. For example, if dichotomic measurements are chosen, the Clauser-Horne-Shimony-Holt (CHSH) inequality~\cite{Clauser1970} can be used. In the following section, we examine this issue in the case of three dichotomic measurements as per party.

As depicted in Fig.~\ref{fig:DISM}, we have established a DI framework to verify EPR-steering. As all pure bipartite entangled states and the associated measurements could be self-tested~\cite{Yang2013,Coladangelo2017}, together with experimental confirmations~\cite{Zhang2019a,Zhang2019b}, it is naturally to witness all steerable states via this DI protocol.

\section{Three measurement settings} 
If Bob receives $\tau^T_{c, j}=(\mathbb{I}+c\sigma_j)/2$ with $c=\pm 1$ and $j=1,2,3$ sent from Charlie where $\sigma_j$ represent three Pauli observables as required in Eq.~(\ref{QRS}), then they can be self-tested if the following triple Bell-CHSH inequality~\cite{SM,Acin2016}
\begin{align}
\mathfrak{B}=&E_{1,1}+E_{2,1}+E_{1,2}-E_{2,2}\nn\\
	&+E_{3,1}+E_{4,1}-E_{3,3}+E_{4,3}\nn\\
	&+E_{5,2}+E_{6,2}-E_{5,3}+E_{6,3}\label{eq:CHSH}
\end{align}
is maximally violated within quantum theory, where $E_{y,z}=\sum _{b,c=\pm 1}b\,c\,p(b,c|y,z)$ refers to the measurement expectations between Bob's dichotomic measurements $y=1,2,...,6$ and Charlie's $z=1, 2, 3$. Specifically, its maximal quantum violation $\mathfrak{B}_{\max}=6\sqrt{2}$ is achieved at Bob's six measurements $(\sigma_i\pm \sigma_j)/\rt{2}$ with $(i,j)=\{(3, 1),(3, 2),(1, 2)\}$ and Charlie's $\sigma_1, \pm\sigma_2,\sigma_3$ on $\ket{\Phi^+_2}=(\ket{00}+\ket{11})/\rt{2}$, up to a local unitary. Note that there is a sign problem in the second measurement $\sigma_2$ for Charlie, however, it does not affect its utility in the DI steering protocol just as the DI entanglement certification~\cite{Bowles2018a}.

Generally, it is impossible to achieve the perfect self-testing with the violation bound $6\rt{2}$. To evaluate imperfections of self-testing, we introduce the fidelity $f_0=\bra{\Phi^+_2}\rho^0_{\rm data}\ket{\Phi^+_2}$ which measures the overlap between a state self-tested from experimental data and the target state. Correspondingly, the fidelity for Charlie's measurements $j=1, 2, 3$ can be cast as the state fidelity in a form of $f_j=\bra{\Phi^+_2}\mathbb{I}^B\otimes\sigma^C_j\rho^j_{\rm data}\mathbb{I}^B\otimes\sigma^C_j\ket{\Phi^+_2}$. All can be computed via a semi-definite program~\cite{Yang2014,Bancal2015,NPA2008}. Incorporating these fidelity into the DI steering inequality~(\ref{DI0}), we obtain 
\begin{equation}
W^{\rm noisy}_{\rm DI}=W_{\rm DI}-\sum_{j=1}^3 \rt{1-f_j}/2\leq 0. \label{noisy}
\end{equation}
Its tedious derivation is deferred to the Supplementary Material~\cite{SM}. This inequality accounts for the imperfect self-testing in terms of state fidelity, interchangeable with the trace distance used~\cite{Bowles2018b,SM}. It also differs from the one in~\cite{Kocsis2015} which is obtained via tomography.

\section{Experimental setup} The experimental setup for DI verification of EPR-steering is displayed in Fig.~\ref{fig:setup}. In particular, two pairs of entangled photons pairs are first generated via the spontaneous parametric down-conversion process. One pair labelled as $\rho_{34}$ in the setup is prepared as the maximally entangled state $\ket{\Phi^+_2}=(\ket{00}+\ket{11})/\rt{2}$ where the horizontally polarised direction (H) and vertically polarised direction (V) encode as state basis $\ket{0}, \ket{1}$, respectively. While, the other pair is generated in a family of Werner states
\begin{align} \label{Werner}
	\rho_{12}=v\ket{\Psi^-_2}\bra{\Psi^-_2}+(1-v)\frac{\mathbb{I}}{4},~~~v\in [0, 1].
\end{align}
Here, \xinhui{$\ket{\Psi^-_2}=\frac{1}{\sqrt{2}}(\ket{01}-\ket{10})$} and the white noise with $1-v$ in Eq.~\eqref{Werner} is simulated by flipping Alice's measurements with probability $(1-v)/2$~\cite{Saunders2017}. This class of states will be tested by the noisy steering witness~\eqref{noisy}.

Then, these photonic states are distributed to three observers. As shown in left side of Fig.~\ref{fig:setup}, $\rho_{12}$ is sent to Alice (photon 1: the green ball)  and Bob (photon 2: the blue ball) through single-mode fibres while $\rho_{34}$ is distributed to Charlie (photon 3: the red ball) and Bob (photon 4: the yellow ball). Detailed rotation parameters adjusted for wave plates (WPs) to realise Alice's and Charlie's three Pauli measurements $\sigma_j$ and Bob's six measurements are given in Tab.I in~\cite{SM}. In the right side of Fig.~\ref{fig:setup}, Bob's partial BSM is implemented via three polarising beam splitters, two $22.5^{\circ}$ rotated HWPs, and four pseudo photon-number-resolving detectors (PPNRD). In each PPNRD, a balanced beam splitter splits the light into two fibre-coupled single photon detector. To improve the quality of the partial BSM, an interference filter of 2 nm is inserted for spectral selection so that a Hong-Ou-Mandel interference visibility higher than 30 : 1 is observed in this experiment. We also do tomography to reconstruct the BSM and obtain a fidelity around $0.9831\pm 0.0040$~\cite{SM}.

Finally, the measurement statistics is collected to do the triple Bell-CHSH test~(\ref{eq:CHSH}) to self-test quantum states $\tau_{c, j}^T=(\mathbb{I}+c\sigma_j)/2$ input to Bob. Combining with the measurement fidelity $f_j$ estimated from imperfect self-testing, we can rewrite the DI steering inequality~(\ref{noisy}) explicitly as~\cite{SM}
\begin{align}
	&4\sum_{j,a,c} \left(a_j\,c_j\,P(a, {\rm Yes}, c\,|x=z=j, \mathcal{B})-P(a, c\,|j)/\rt{3}\right)\nn\\
	-&\xinhui{2}\sum_j\rt{1-f_j}\leq 0\label{payoff function}.
\end{align}
\xinhui{Here $g_{c, j}$ is either $1$ or $-1$ for the qubit measurements}. For Werner states~(\ref{Werner}), the theoretical prediction of Eq.~(\ref{payoff function}) should be $3v-\rt{3}-2\sum_{j=1}^3\rt{1-f_j}\xinhui{\leq 0}$. It is found that the average fidelity around  \xinhui{$99.7\%$} of self-testing is allowed for Bell local states with $v=0.7$~\cite{SM}, which is a significant reduction in comparison to entanglement verification with fidelity above $99.998\%$~\cite{Bowles2018b}.

\section{Results}
The entangled photon pairs encoding $\ket{\Phi^+_2}$ are collected up to $13,000$ per second with a pump power of $30$~mW. We observe an extinction ratio over $500:1$ in the H/V basis and the H+V/H-V basis, indicating it is generated with fidelity higher than $0.998$. Under the fair-sampling assumption, we obtain $2.8241$, $2.8211$, and $2.8189$ for three Bell-CHSH tests in~(\ref{eq:CHSH}) and the sum of them is 8.4641 closing to the maximal quantum bound $6\sqrt{2}\approx 8.4853$. The uncertainty induced by the Poisson oscillation of photons is about 0.0009. Correspondingly, the fidelity of three Pauli measurements self-tested from experimental data is \xinhui{$f_1=0.9994, f_2=0.9999,$ and $f_3=0.9992$}, respectively, and thus the average fidelity $99.95\%$ is attained in our experiment. The standard deviation is around $10^{-5}$ by optimising 100 groups of the Poisson statistics of the experimental  data.

\begin{figure}[ht]
	\centering
	\includegraphics[width=1.0\linewidth]{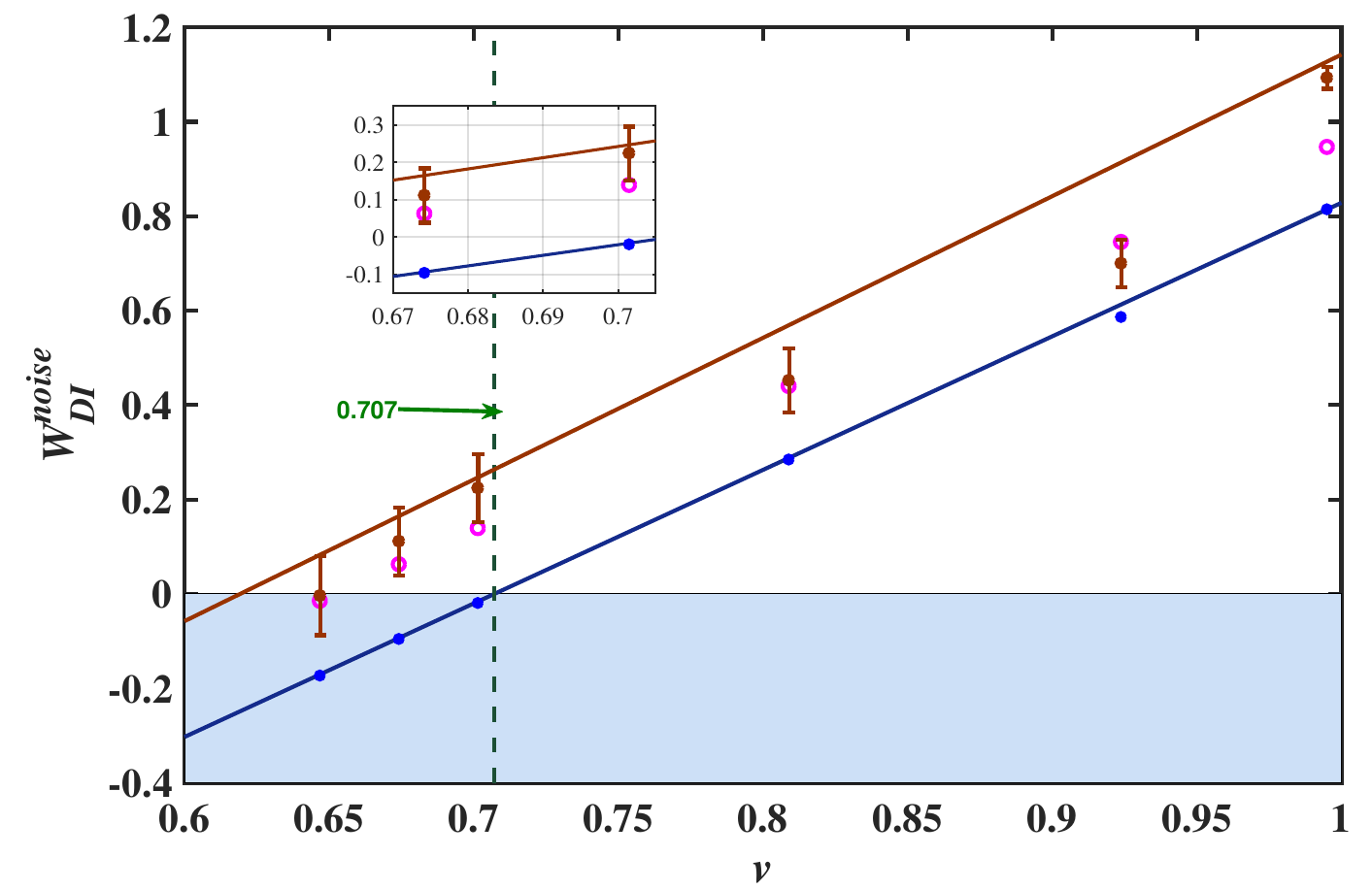}
	\caption{{\bf Experimental results.} The theoretical prediction for ideally DI  verification~(\ref{payoff function}) is plotted in the red curve, while  the experimental results are in red dots for $v=0.6469(4), 0.6742(4), 0.7015(4), 0.8090(4), 0.9239(3),0.9951(1)$. We observe a violation up to $0.1110\pm0.07$ even for a Bell local state with $v=0.6742(4)$, smaller than the bound $v=0.707$ for the CHSH inequality. \xinhui{For the reason of the small amount of data, about three orders of magnitude small,  the error bars for DI verification (\ref{payoff function}) are much larger than the CHSH inequality. To make a comparison with fully DI, we do tomography on entangled sources and partial BSM and obtain a post-processing probability statistics in DI steering (\ref{payoff function}) where the fidelity of local measurements are estimated by self-testing, the results are shown as the magenta circles.} By contrast, the corresponding theoretical and experimental results for the \xinhui{standard} CHSH inequality with two measurements \xinhui{per party are given in blue curve and blue dots respectively, where the bounds are translated down by 2}. The error bars are about $0.001$. The shaded blue region represents the failure of the DI steering inequality. }
	\label{fig:result}
	\rule[-10pt]{8.5cm}{0.05em}
\end{figure}

The experimental results of DI verification for Werner states are plotted in Fig.~\ref{fig:result}. We perform quantum state tomography to show that each Werner state is generated with $v=0.6469(4)$, $0.6742(4)$, $0.7015(4)$, $0.8090(4)$, $0.9239(3)$, and $0.9951(1)$ from about $10^7$ photon pairs~\cite{SM}. Ideally, there is the theoretical prediction $3v-\rt{3}$ for $f_j=1$, yielding the EPR-steering bound $v=1/\sqrt{3}\approx0.5774$~\cite{Cavalcanti2009}. Otherwise, the steering inequality~(\ref{payoff function}), incorporated with self-testing imperfections, is shown as the red line in Fig.~\ref{fig:result}, while the experimental results are displayed in red dots. It is evident that we have successfully witnessed steerability device-independently if $v\geq0.6742(4)$, accounting for statistic errors and imperfections of self-testing. Importantly, a violation up to $0.1110\pm0.07$ is achieved at the point $v=0.6742(4)$ lower than the CHSH bound $1/\sqrt{2}\approx 0.707$~\cite{Clauser1970} and the V\'etersi bound $\gtrsim 0.7056$~\cite{Vertesi2008}. This implies that even some Bell local states can be verified via this DI protocol. The error bars for Werner states with $v=0.6469(4)$ fall into the failure region, so their steerability are not conclusively detected. 

We further test the system errors on the performance of DI steering verification. We do quantum tomography on the entangled sources and partial BSM. This process is given in~\cite{SM}, and the calibrated violations of DI steering inequality~(\ref{payoff function}) are shown as the magenta circles in Fig.~\ref{fig:result}. By contrast, we also perform the CHSH test to verify steerability. In Fig.~\ref{fig:result}, the blue line describes the theoretical results while blue dots are the experimental observations for these Werner states.

\section{Conclusion and discussion} We have studied the DI verification of EPR steering and implemented an optical experiment to validate our DI protocol. In principle, we prove that all steerable states, including Bell local states, can be verified device-independently. In practice, we analyse noise robustness towards imperfections of self-testing in the implementation process, and derive a steering inequality as per Eq.~(\ref{noisy}) for the three-measurement setting case. Finally, we perform a proof of principle experiment to successfully validate our DI steering protocol. We believe that our work paves the way for realistic implementations of secure quantum information processing tasks based on EPR-steering and also finds practical applications of self-testing.

There are many interesting open questions left for the future work. For example, the methods in~\cite{Bennet2012,Rutkowski2017} may be used to tolerate more transmission loss and lower measurement efficiency. The resource efficient approach in~\cite{Verbanis2016} could also improve the success probability of the partial BSM, and self-testing could be more noise robust by adopting other techniques~\cite{Supic2019}. Moreover, an alternate DI framework~\cite{Hall2019} may be possibly used to verify quantum steering. It is also interesting to device-independently certify genuine high-dimensional steering~\cite{Designolle2021} and steering networks~\cite{Jones2021}.

\section*{Acknowledgments}
We acknowledge V. Scarani, M. J. W. Hall, Y. Han, and Y. Wang for constructive comments.
 This work  was supported by  the National Natural Science Foundation for the Youth of China (No.11704371), China Postdoctoral Science Foundation (No. 2017M612074, BH2030000088), the National Key Research and Development Program of China (No. 2017YFA0304100, 2016YFA0301300, 2016YFA0301700), NSFC (No. 11774335, 11734015, 11504253, 11874345, 11821404, 11774334, 12134014), the Key Research Program of Frontier Sciences, CAS (No. QYZDY-SSWSLH003), Science Foundation of the CAS (ZDRWXH-2019-1), the Fundamental Research Funds for the Central Universities, USTC Tang Scholarship, and Anhui Initiative in Quantum Information Technologies (No. AHY020100, AHY060300, 2008085J02); \xinhui{Y. Zhao is supported  by the Major Key Project of PCL.} X. Li and Q. Wen are supported by NSFC (No. 61672110, 61671082, 61976024, 61972048) and the Fundamental Research Funds for the Central Universities (No. 2019XD-A01); \xinhui{S. L. Chen and H. Y. Ku are supported by National Science and Technology Council, Taiwan (Grants Nos. MOST 111-2917-I-564-005 and 111-2112-M-005-007-MY4).}

\newpage

\hspace{10cm}

\newpage

\appendix

\onecolumngrid

\vspace*{0.4cm}
\begin{center}
	{\large \bf Supplemental material for:}
	\vskip 0.2 cm
	{\large \bf Device-independent verification of Einstein-Podolsky-Rosen steering} \\
	\vspace{0.6cm}

    \vspace{0.1cm}
\vspace{0.6cm}
\end{center}
\onecolumngrid

\setcounter{equation}{0}
In this supplementary material, we give a detailed analysis of fully device-independent (DI) verification of Einstein-Podolsky-Rosen (EPR) steering or quantum steering step by step. First, the standard EPR-steering is introduced and its detection is discussed. Then, we move to measurement-device independent (MDI) verification of EPR-steering, an important step to eliminate the trust in measurement devices with additional assumptions. Further, by using self-testing to remedy above extra assumptions, we arrive at a fully device-independent (DI) verification protocol. Moreover, the noise robustness of our DI steering protocol is analysed, especially robustness of self-testing, and a DI steering inequality is constructed to expose steerability of physical states, which naturally certifies the presence of entanglement within quantum theory.  Finally, the optical experimental details to implement the complete DI verification of EPR-steering are presented.

\section{What is EPR-steering?}

Suppose that two observers, namely Alice and Bob, make some measurements on a preshared state (they may not have a quantum description). Steering was first introduced by \sch ~to describe the ability that Alice's local measurements could prepare Bob's states remotely~\cite{schrodinger1935}, and this phenomenon was generalised as EPR -steering by Wiseman {\it et al.}~\cite{Wiseman2007}. If all follows quantum rules, it has an operational interpretation as an entanglement verification task. Specifically, if Alice's and Bob's measurements are labeled as $x$ and $y$ respectively, and the corresponding outcomes $a$ and $b$, this task amounts to checking if the collected statistics $p(a,b|x,y)$ admit a {\it local hidden state} (LHS) model in a form of
\beq
p(a, b|x, y)=\sum_\lambda p(\lambda) p(a|x, \lambda) {\rm Tr}[E^B_{b|y}\rho^B_\lambda], \label{LHS}
\eeq
where the hidden variable $\lambda$ specifies some classical probability distribution $p(a|x, \lambda)$ for Alice and some quantum probability distribution ${\rm Tr}[E^B_{b|y}\rho^B_\lambda]$ for Bob which is generated via performing a positive-operator-valued measurement (POVM) $\{E^B_{b|y}\}_{b, y}$ on quantum states $\rho^B_\lambda$~\cite{Wiseman2007}. If there is no such LHS model, then EPR-steering from Alice to Bob is demonstrated.

In principle, every steerable state can be witnessed in an experimental-friendly manner by violating a suitable linear steering inequality of the form~\cite{Cavalcanti2009}
\beq
W_{\rm S}=\sum_{j}\an{a_jB_j}\leq 0, \label{steerineq}
\eeq
where $a_j$ represents the outcome of Alice's measurement $j$ and Bob's correlated measurement $j$ has a quantum-meachnical description $B_j$. For example, consider the measurement scenario where Alice and Bob are specified to three dichotomic measurements. If Bob's measurements are further chosen as mutually unbiased observables, it immediately gives rise to a steering inequality~\cite{Cavalcanti2009}
\begin{align}
	W_{\rm S}&=\an{a_0\sigma_0+a_1\sigma_1+a_2\sigma_2+a_3\sigma_3} \nn \\
	&=\an{a_1\sigma_1+a_2\sigma_2+a_3\sigma_3}-\rt{3}\leq 0,  \label{optimalwitness}
\end{align}
where $\sigma_0=\mathbb{I}$, $a_0=-\rt{3}$, and operators $\sigma_j$ for $j=1, 2, 3$ correspond to three Pauli operators $\sigma_\mathrm{x}, \sigma_\mathrm{y}, \sigma_\mathrm{z}$. With respect to the family of Werner states in the main text,
\beq
\rho=v\ket{\Psi^-_2}\bra{\Psi^-_2}+(1-v)\frac{\mathbb{I}}{4},~~~v\in [0, 1] \label{ApWerner}
\eeq
with $\ket{\Psi_2^-}=\frac{1}{\rt{2}}\left(\ket{01}-\ket{10}\right)$, it is easy to check that $W_{\rm S}(\rho)=3v-\rt{3}$. So, if the visibility is larger than the bound $\rt{3}/3\approx 0.577$, i.e., violating this steering inequality, then the steerability of this class of states is exposed.

\section{How can we verify EPR- steering measurement device-independently?}

Given the measurement outcome statistics $p(a, b|x, y)$ in Eq.~(\ref{LHS}), if Alice's side also admits a quantum description, then the above task reduces to the entanglement verification. In a seminal work~\cite{Buscemi2012}, Buscemi established a MDI framework to certify all entangled states, in which neither Alice nor Bob is trusted or assumed to follow quantum rules. Indeed, the trust in both sides is completely transferred to a third observer, Charlie say, who could prepare a set of quantum states and then randomly assigns them to either Alice or Bob.

The MDI framework was later extended to EPR-steering~\cite{Cavalcanti2013}. Since Alice is already device-independent, Bob's trust is the only issue to be addressed. In the MDI scenario, Bob and his device are not trusted any more, and thus the quantum probability for Bob in Eq.~(\ref{LHS}) and the steering inequality with $B_j$ as per Eq.~(\ref{steerineq}) are not applicable neither. It works that Bob is instead specified to a set of quantum states $\{\tau^T_{b, j} \}$ at random from Charlie where $T$ is the transpose operation. Then, Bob is required to perform some joint measurement $\mathcal{B}$ on his subsystem and the input quantum states and reply with dichotomic outputs denoted by ``{\rm No}'' and ``{\rm Yes}'', respectively. Denote by $P(a,{\rm Yes}\,|x, \mathcal{B}, \tau^T_{b,j})$ the probability that Alice obtains $a$ for the measurement $x$ and Bob answers ``Yes'' when assigned to  $\tau^T_{b,j}$. It is possible that arranging these outcome statistics properly yields a quantum-refereed steering (QRS) witness~\cite{Kocsis2015,Guo2018}
\beq\label{eq:WQRS1}
W_{\rm QRS}=\sum_{j, a, b} g_{b,j}a_j\,P(a, {\rm Yes}\,|x=j,\mathcal{B},\tau^T_{b, j})\leq 0,
\eeq
where $g_{b, j}$ are some predetermined parameters. In practice, Bob could perform a partial Bell state measurement (BSM) $\mathcal{B}\equiv\{\mathcal{B}_1, \mathbb{I}-\mathcal{B}_1\}$ where $\mathcal{B}_1 = | \Phi^+_d\rangle\langle \Phi^+_d|$ with $| \Phi^+_d\rangle=\sum_j|jj\rangle/\rt{d}$ models the answer ``Yes'' and $d$ is the dimension of the Hilbert space of $\{\tau^T_{b,j}\}$ equal to that of Bob's local system. Note that Bob's observables $B_j$ in Eq.~(\ref{steerineq}) could be decomposed into a linear combination of their outcomes which are modelled by elements $E_{b|j}$ of POVMs, i.e., there is
\beq
B_j=\sum_b g^\prime_{b, j}E_{b|j},~~E_{b|j}\geq 0,~~\sum_bE_{b|j}=\mathbb{I},
\eeq
where $b$ refers to the measurement outcome of $B_j$. If Alice and Bob share a state $\rho_{AB}$ to be tested, then the above QRS becomes
\begin{align}\label{eq:WQRS2}
	W_{\rm QRS}&=\sum_{j, a, b}a_j\,g_{b,j}\tr{\xinhui{E_{a|j}}\otimes \mathcal{B}\cdot \rho_{AB}\otimes \tau^T_{b, j}}=\frac{1}{d}\sum_{j, a, b}a_j\,g_{b,j}\tr{\xinhui{E_{a|j}}\otimes \tau_{b, j}\cdot \rho_{AB}}.
\end{align}
\xinhui{Here the measurement element $E_{a|j}$ describes Alice's measurement outcome $a$ given a measurement $j$.} When these input states are chosen as $\tau_{b, j}=E_{b|j}$ and the predetermined parameters satisfy $g_{b, j}=g^\prime_{b, j}$, it leads to
\begin{align}
	W_{\rm QRS}=\frac{1}{d}\sum_{a,j} a_j\tr{\xinhui{E_{a|j} }\otimes B_j\rho_{AB}}=\frac{1}{d}\sum_j\an{\xinhui{A_j\otimes B_j}}=\frac{1}{d}W_{\rm S},
\end{align}
as in the main text, \xinhui{with $ A_j\equiv \sum_a a_jE_{a|j}$} . It was shown in~\cite{Kocsis2015,Guo2018} that each QRS witness can be constructed from a standard steering inequality as per Eq.~(\ref{steerineq}), implying that all steerable states can be witnessed in an MDI manner.

For the class of Werner states given in Eq.~(\ref{ApWerner}), when Bob is randomly input to
\begin{align}
	\tau_{b,j}=\frac{1}{2}\left(\mathbb{I}+b\sigma_j \right), b=\pm 1 \label{outcome}
\end{align}
with $g_{b, j}=b=\pm 1$ and performs a partial BSM, it is easy to derive that $W_{\rm QRS}(\rho)=(3v-\rt{3})/2$.

\section{How can we verify EPR-steering device independently?}

It follows from above discussions that in the MDI framework both Alice's and Bob's side are already device-independent or trust-free, while the extra trust in the preparation of quantum states $\{\tau^T_{b, j} \}$ by Charlie is still required. Hence, eliminating this trust in Charlie immediately gives rise to a fully DI steering verification. One possible way to addressing this issue is self-testing \cite{Supic2019} which refers to a device-independent way to uniquely identify the state and the measurement for uncharacterised quantum devices. As the only information required is the number of measurements, the number of outputs of each measurement, and the outcome statistics, it is thus a completely device-independent process.

As discussed in the main text, in the fully DI verification of EPR-steering framework, we need to collect the measurement statistics to check whether it violates the DI steering inequality
\beq
W_{\rm DI}=\sum_{a,c,j} g_{c,j} a_j P(a, {\rm Yes}, c\,|x=j, \mathcal{B}, z=j)\leq 0 \label{DI}
\eeq
for the ideal case, given any quantum state $\rho_{AB}$ to be tested. For example, suppose that Bob is input $\tau^T_{b,j}=\frac{1}{2}\left(\mathbb{I}+b\sigma_j \right)$ with $b=\pm 1, j=1,2,3$ randomly from Charlie. Alternate, Bob's input states could be generated by Charlie performing local measurements described by $\{E_{b, j}=\tau_{b,j}\}$ on the Bell state $\ket{\Phi^+_d}$ shared by Bob and Charlie. Thus, the DI steering inequality~(\ref{DI}) could be expressed in a more explicit form of
\begin{align}\label{eq:QRS}
	W_{\rm DI}=&\sum_{a,b,j} b_{j} a_j P(a, {\rm Yes}, c\,|x=j, \mathcal{B}, z=j)\nn\\
	=&\sum_{j, a, b}a_j\,b_{j}{\rm Tr}[\xinhui{E_{a|j}}\otimes \ket{\Phi_2^+}\bra{\Phi^+_2}_{BB_0}
	\otimes (\mathbb{I} +b\sigma^C_j)/2\cdot \rho_{AB}\otimes \ket{\Phi_2^+}\bra{\Phi^+_2}_{B_0C}] \nn \\
	=&\frac{1}{4}\sum_{j, a, b}a_j\,b_{j}\tr{\xinhui{E_{a|j}}\otimes(\mathbb{I}+b\sigma^B_j)/2 \cdot \rho_{AB}}\nn\\
	=&\frac{1}{4}\sum_j\an{\xinhui{A_j} B_j}=\frac{1}{2}W_{\rm MDI}=\frac{1}{4}W_{\rm S}.
\end{align}
Here $B_0$ represents the Bob's subsystem that  $\tau_{b,j}=\frac{1}{2}(\mathbb{I}+b\sigma_j)$.

In particular, these trust input states  $\tau_{b,j}=\frac{1}{2}(\mathbb{I}+b\sigma_j)$ 
for Bob in MDI steering scenario can be replaced by these untrusted observables
via self-testing which refers to a device-independent way to uniquely identify the
state and the measurement for uncharacterized quantum
devices. The virtual protocol that one considers is described as following.

Consider the scenario in which involves two
non-communicating parties Bob and Charlie. Each has access to a black box
with an underlying state $\ket{\psi}$.
It is accomplished with three Bell-CHSH tests and thus Bob needs to perform the six dichotomic measurements $y=1,2,...,6$ and Charlie performs $z=1,2,3$. 
Bob's inputs and outputs are denoted respectively by $y$ and $b$;
Charlie's by $z$ and $c$. After a large number of rounds of experiments, the joint probability distribution $p(b,c|y,z)$ could be reconstructed. Then we are able to construct the triple Bell operator defined in ref.~\cite{Acin2016}
\begin{subequations}
\begin{align}
	\mathfrak{B}=&E_{1,1}+E_{2,1}+E_{1,2}-E_{2,2}\label{eq:CHSHa}\\
	&+E_{3,1}+E_{4,1}-E_{3,3}+E_{4,3}\label{eq:CHSHb}\\
	&+E_{5,2}+E_{6,2}-E_{5,3}+E_{6,3}\label{eq:CHSHc}.
\end{align}
\end{subequations}

\begin{figure}[ht]
\centering
\includegraphics[scale=0.4]{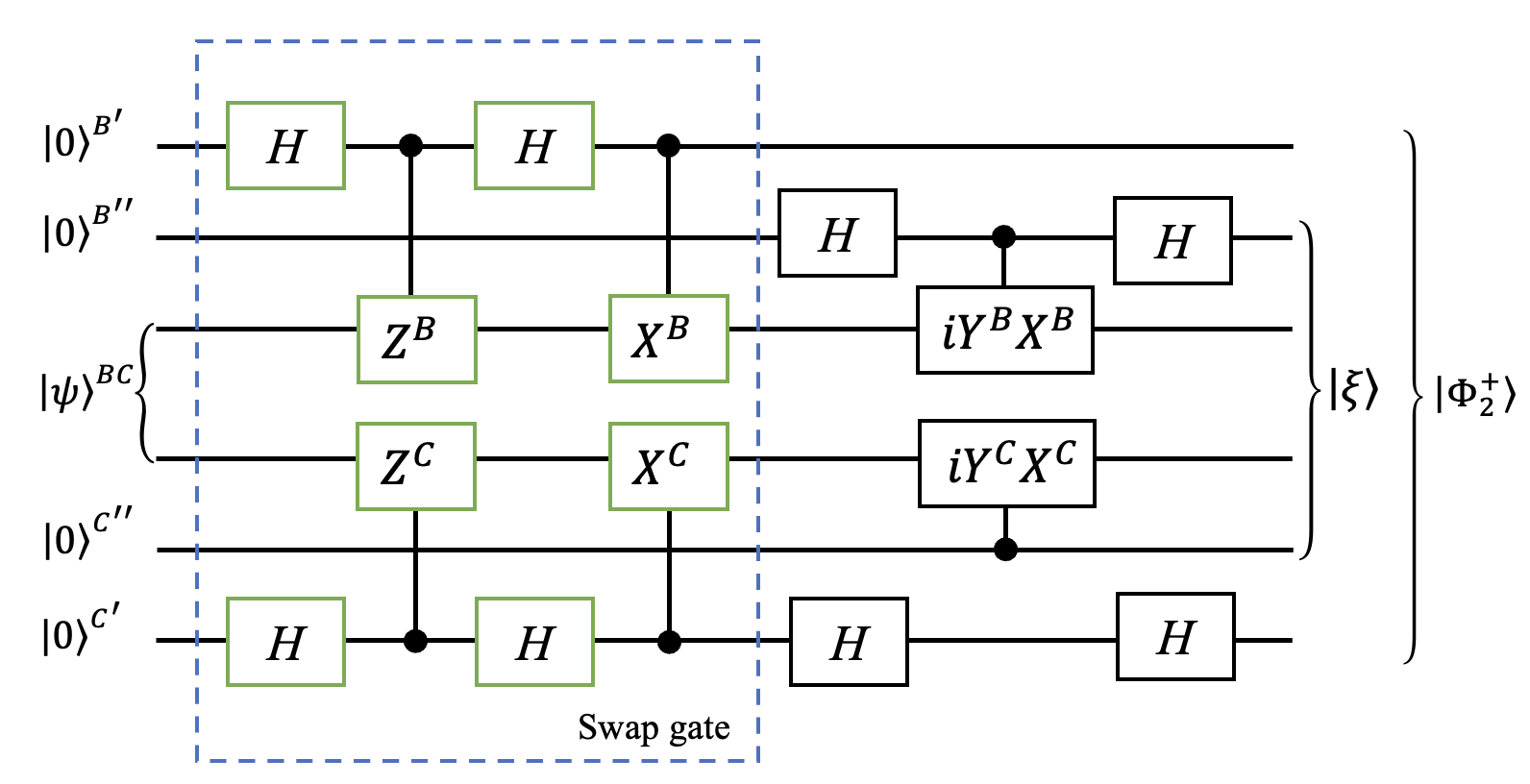}
\caption{\xinhui{The local isometry $U$ is explicitly constructed 
	to self-test the singlet state and Pauli operators. The isometry is a virtual protocol, all that must be done in laboratory is to query the boxes and 
derive $p(b,c|y,z)$.}}
\label{fig:Isometry}
\rule[-10pt]{13cm}{0.05em}
\end{figure}

Further, it was proven by Bowles {\it et al.}~\cite{Bowles2018b} that if the maximal quantum violation $\mathfrak{B} =6\sqrt{2}$ is observed, then there exists a local auxiliary state $\ket{00}\in[\mathcal{H}_{B'}\otimes\mathcal{H}_{B''}] \otimes[\mathcal{H}_{C'}\otimes\mathcal{H}_{C''}]$ ($\ket{00}$ is short for $\ket{0000}_{B'B''C'C''}$) and a local isometry $U$ (see Fig.~\ref{fig:Isometry}) such that
\begin{align}\label{eq:self-test M}
    U[M_i^C\ket{\psi}\otimes\ket{00}]&=\ket{\xi}\otimes \sigma_i^{C'}\ket{\Phi^+_2}^{B'C'},\nn\\
    U[Y^C\ket{\psi}\otimes\ket{00}]&=\sigma^{C''}_\mathrm{z}\ket{\xi}\otimes\sigma^{C'}_\mathrm{y}\ket{\Phi^+_2}^{B'C'},
\end{align}
where $M_i\in\{I,X,Z\}$, $\sigma_i\in\{I,\sigma_x,\sigma_z\}$ and $\ket{\xi}$ is the junk state left in systems $[\mathcal{H}_{B}\otimes\mathcal{H}_B'']\otimes[\mathcal{H}_{C}\otimes\mathcal{H}_C'']$, in the form of
\begin{align}\label{st:junk}
    \ket{\xi}=\ket{\xi_0}^{BC}\otimes\ket{00}^{B''C''}+\ket{\xi_1}^{BC}\otimes\ket{11}^{B''C''}
\end{align}
with $\langle\xi_0|\xi_0\rangle+\langle\xi_1|\xi_1\rangle=1$. It means that we can extract the exact information of the maximally entangled state of two-qubit $\ket{\Phi_2^+}=\frac{1}{\sqrt{2}}(\ket{00}+\ket{11})$ and Charlie's three measurements
\begin{align}
X^C=\sigma_\mathrm{x},\;\;Y^C=\pm \sigma_\mathrm{y}  \;\;Z^C=\sigma_\mathrm{z}
\end{align}
Although there exists the sign problem of $\sigma_y$ to be distinguished, it does not pose any constraint to verify entanglement~\cite{Bowles2018a} and EPR steering to be discussed.

Note that the measurement set $\{\sigma_x, -\sigma_y, \sigma_z\}$ could be transformed from the set $\{\sigma_x, \sigma_y, \sigma_z\}$ on which is acted the transpose operation $T$ because of $\sigma_y^T=-\sigma_y$. It is easy to verify that the state $\rho_{AB}$ has a local hidden state (LHS) model with respect to one measurement if and only if it holds for the other measurement set, since the partial operation does not change the nonlocal property of EPR steering. Thus, the sign of $\sigma_y^C$ is not a problem when we certify the steerability. In regard to Werner states $\rho_{12}=v\ket{\Psi^-_2}\bra{\Psi^-_2}+(1-v)\frac{\mathbb{\mathbb{I}}}{4},~v\in [0, 1]$, we have
\beq
W_{\rm DI}=\frac{1}{4}(3\,v-\rt{3}).
\eeq
This again indicates that we can verify all steerable states with an DI protocol.

Hence, we can obtain a DI protocol, combining MDI techniques with self-testing, to verify every steerable state. 

\section{Robust DI verification of EPR-steering }

Ideally, our results derived work well. However, due to imperfections, such as transmission loss or measurement errors, we may collect the noisy data which is usually unable to violate the Bell inequality maximally. Thus the self-testing process is not perfect, and we need to estimate the distance between the observed statistics and the targeted one, a property known as \textit{robustness}. In this section, we give a detailed analysis of robust self-testing for Pauli observables based on Navascu\'es-Pironio-Ac\'in (NPA) hierarchy and the semi-definite program (SDP). Then, we provide an DI steering inequality, allowing for imperfections of self-testing.

\subsection{Robust self-testing of Pauli observables}

In the ideal case, we have constructed a local isometry to certify the two-qubit Bell state $\ket{\Phi^+}$ from the unknown physical state $\ket{\psi}^{BC}$. Similarly, three Pauli observables $\sigma_j, j=1,2,3$ are cast as the state self-testing of $\sigma^C_j \ket{\Phi^+}$ from uncharacterised $M^C\ket{\psi}$, where $M^C\in\{X,Y,Z\}$ is the unknown local operator acting on Charlie. As shown in Fig.~\ref{fig:Isometry}, this isometry circuit is a swap circuit~\cite{Yang2014} composed of a set of controlled gates and Hadamard gates. The idea of the swap method is to ``swap'' out the essential information onto auxiliary systems with the same dimensionality as the local systems of the target state.

\xinhui{We first consider the state  $M^C\ket{\psi}^{BC}$ and local auxiliary state $\ket{00}\in \mathcal{H}_{B''}\otimes\mathcal{H}_{C''}$ through the swap gate part in the circuit shown in Fig. \ref{fig:Isometry}, which becomes 
\begin{align} 
U_{swap}M^C\ket{\psi}^{BC}\ket{00}^{B''C''}=&\frac{1}{4}
[(\mathbb{I}+Z_B)(\mathbb{I}+Z_C)M^C\ket{\psi}\ket{00}
+X_C(\mathbb{I}+Z_B)(\mathbb{I}-Z_C)M^C\ket{\psi}\ket{01}\nn\\
&+X_B(\mathbb{I}-Z_B)(\mathbb{I}+Z_C)M^C\ket{\psi}\ket{10}
+X_B(\mathbb{I}-Z_B)X_C(\mathbb{I}-Z_C)M^C\ket{\psi}\ket{11}.
\end{align}
By denoting $\ket{\phi}=U_{swap}M^C\ket{\psi}^{BC}\ket{00}^{B''C''}$, then two Hadamard gates combing with the third pair of controlled gates evolves the system to be
\begin{align}
U[M^C\ket{\psi}^{BC}\ket{00}^{B'C'}\ket{00}^{B''C''}]=&\frac{1}{4}(\ket{++}^{B''C''}\ket{\phi}+\ket{+-}^{B''C''}iY_CX_C\ket{\phi}\nn\\
&+\ket{-+}^{B''C''}iY_BX_B\ket{\phi}-\ket{--}^{B''C''}Y_BX_BY_CX_C\ket{\phi}),
\end{align}
where $\ket{\pm}=\ket{0}\pm \ket{1}$.} To extract the information of the trusted auxiliary systems $B'$ and $C'$, we take the partial trace of the whole system which be left 
\begin{align}
\rho^j_{\rm data}={\rm Tr}_{BB''CC''}(U\rho^j_{BC}\ket{0000}\bra{0000}_{B'B''C'C''}U^\dagger)=\frac{1}{64}\sum\limits_{m,n,k,l\in\{0,1\}}C^j_{mnkl}\ket{m}\bra{n}\otimes\ket{k}\bra{l}\label{st:swap},
\end{align}
where  $\rho^j_{BC}=M^C_j\ket{\psi}\bra{\psi}{M^C_j}^\dagger$ describes the density matrix of untrusted operator $M^C_j$ acting on the uncharacterised state $\ket{\psi}$ and $C^j$ is the coefficient matrix of $\rho^j_\text{data}$ with 
\begin{align}
C^j_{mnkl}=&{\rm Tr}_{BB''CC''}[{(-iY_BX_B)}^{m} (I+Z_B)^{1-m}(X_B-Z_BX_B)^m(iY_BX_B)^n(I+Z_B)^{1-n}(X_B-X_BZ_B)^n\nn \\
&\otimes {(-i X_CY_C)}^{k}(I+Z_C)^{1-k}(X_C-Z_C X_C)^k(iY_CX_C)^l(I+Z_C)^{1-l}(X_C-X_CZ_C)^l  \rho^j_{BC}].
\end{align}
\xinhui{Looking into these single terms, it can be found that for each target Pauli observable, $\rho_{data}$ is a $4 \times 4$ matrix whose entries are linear combinations of expectation values such as $\langle X_A \rangle$, $\langle X_AZ_A \rangle$, $\langle X_AZ_B \rangle$, $\langle X_AZ_BX_C \rangle$, etc. Then the closeness of $\rho^j_{\rm data}$ to the target state $\sigma^{C}_j\ket{\Phi^+_2}$ can be then captured by the fidelity
\begin{align}
f_j=\bra{\Phi^+_2}\mathbb{I}^{B'}\otimes\sigma^{C'}_j \rho^j_{\rm data}\mathbb{I}^{B'}\otimes\sigma^{C'}_j\ket{\Phi_2^+}, j=1, 2, 3.
\end{align}
Here $f_j$ is a linear function of two types of operator expectations: some observed behavior and some non-observable correlations which 
involve different measurements on the same party which are left as variables. }We define an average fidelity
\begin{align}
\widehat{f}=\frac{1}{3}\sum\limits_{j=1,2,3}f_j
\end{align}
to evaluate the performance of self-testing. 
It is worth noting that $\sigma_\mathrm{y}$ and $-\sigma_\mathrm{y}$ have the same fidelity function and thus $\widehat{f}$ for two measurement settings $\{\sigma_x,\sigma_y,\sigma_z\}$ and $\{\sigma_x,-\sigma_y,\sigma_z\}$ are identical.

Finally, the fidelity $f_j, j=1, 2,3$ are calculated with the aid of the NPA hierarchy characterization of the quantum behaviors~\cite{Yang2014,NPA2008,Bancal2015}, and their lower bound can be computed via a SDP:
\begin{align}\label{eq:MSDP}
\min \quad & \widehat{f} \nonumber\\
\text{s.t.}\quad  &\Gamma \geq 0, \nn\\
&\text{the CHSH operators}  \\
&\eqref{eq:CHSHa}=2.8241, \eqref{eq:CHSHb}=2.8211, \eqref{eq:CHSHc}=2.8189, \nonumber
\end{align}
\xinhui{where $\Gamma$ is so-called NPA moment matrix whose rows and columns are numbered by products belonging to $Q_l$, i.e., $\Gamma_{ij}=\bra{\psi} {Q^i_l}^\dagger Q^j_l\ket{\psi}$, and $Q_l$ is the set of product of $B_y$ and $C_z$ and defined as outer approximations of the quantum set (the level of the hierarchy $l$ is the number of measurements in the product). In our problem, the moment matrix corresponding to $Q_2$, that is to say the products set is with at most operators per party. To improve the precision of fidelity, we 
increased the size of the $\Gamma$ matrix by adding terms such as $\langle A_1A_2A_1\rangle$, $\langle A_2A_1B_1\rangle $, $\langle A_3B_1B_2\rangle $, $\langle A_3A_2A_1A_3\rangle$, $\langle A_2A_1A_2A_1A_2\rangle$, et.al. to contain all the average values $\langle \cdot\rangle$ that appear in the expression of fidelity. It results in the $\Gamma$ matrix having a size of $101\times101$ whose elements are divided into two kinds that observed behavior variables are real and non-observable variables are complex. The total number of constrains is $K=2167$ (28 variables are real and the left are complex). We used the MATLAB modeling language YALMIP and MOSEK as a solver to solve the SDP.} According to our experimental results about the violation of the triple Bell-CHSH test, we obtain the average fidelity \xinhui{$\widehat{f}=0.9995$ and $f_1=0.9994$, $f_2=0.9999$, $f_3=0.9992$} for each Pauli observable.

\subsection{Robust verification of EPR steering}
It easily follows from Eq. \eqref{st:swap} that the distance between the pure state estimated from the experimental data via a SDP and the target state satisfies
\begin{align}\label{eq:norm2fid}
\xinhui{\|U[M^C_j\ket{\psi}_{BC}\otimes\ket{00}]-\ket{\xi}\otimes\sigma^{C'}_j\ket{\Phi^+_2}_{B'C'}||=\sqrt{1-f_j}},
\end{align}
where $\ket{00}\in[\mathcal{H}_{B'}\otimes\mathcal{H}_{B''}]\otimes[\mathcal{H}_{C'}\otimes\mathcal{H}_{C''}]$,  $\ket{\xi}$ is defined as in the form of
\begin{align}\label{st:junk}
    \ket{\xi}=\ket{\xi_0}^{BC}\otimes\ket{00}^{B''C''}+\ket{\xi_1}^{BC}\otimes\ket{11}^{B''C''}
\end{align}
with $\langle\xi_0|\xi_0\rangle+\langle\xi_1|\xi_1\rangle=1$ and \xinhui{$||\bullet||$ denotes the trace distance}. Thus, these fidelity of Pauli observables $f_j$ and the Bell state $f_0$ give us the error estimate when we use the experimental data to do the verification task.

Further, the self-tested pure states in Eq.~(\ref{eq:norm2fid}) could be decomposed as
\begin{align}\label{eq:meas2norm}
U[M^C_j\ket{\psi}\otimes\ket{00}]= \ket{\xi}\otimes \left(\alpha_j \sigma^{C'}_j \ket{\Phi_2^+}+\sqrt{(1-\alpha^2_j)} \ket{\phi^\perp_j}\right).
\end{align}
Here the state vector $\ket{\phi^\perp_j}$ is orthogonal to $\sigma_j^{C'}\ket{\Phi_2^+}$  and it is easy to check that $\alpha_j=\sqrt{f_j}$. For each Pauli observable $\sigma_j$, the deviation from the density matrices output from the swap circuit is
\begin{align}
\Delta_j=&{\rm Tr}_{BB''CC''}\big(U[M^C_j\ket{\psi}\bra{\psi}M^C_j\otimes\ket{00}\bra{00}]U^\dagger\big)-{\rm Tr}_{BB''CC''}\big(\ket{\xi}\bra{\xi}
\otimes\sigma^{C'}_j\ket{\Phi_2^+}\bra{\Phi_2^+}_{B'C'}\sigma^{C'}_j\big)\nn\\
=&(\alpha_j^2-1) \sigma^{C'}_j\ket{\Phi_2^+}\bra{\Phi_2^+}\sigma^{C'}_j
+\alpha_j\sqrt{1-\alpha^2_j}\sigma^{C'}_j\ket{\Phi_2^+}\bra{\phi^\perp_j}\nn\\
&+\alpha_j\sqrt{1-\alpha_j^2}\ket{\phi^\perp_j}\bra{\Phi_2^+}\sigma^{C'}_j
+(1-\alpha_j^2)\ket{\phi^\perp_j}\bra{\phi^\perp_j}.
\end{align}
This matrix has two eigenvalues $\lambda_j=\pm\sqrt{1-\alpha^2_j}=\pm\sqrt{1-f_j}$ by solving the following matrix
\begin{align}
\Delta_j=\left[
\begin{matrix}
\alpha_j^2-1 & \alpha\sqrt{1-\alpha_j^2}\\
\alpha_j\sqrt{1-\alpha_j^2}& 1-\alpha_j^2
\end{matrix}
\right]
\end{align}
in the basis of $\{ \sigma^{C'}_j\ket{\Phi_2^+}, \ket{\phi^\perp_j}\}$. Instead of Charlie's local measurements $\sigma_j$ for the ideal case, $\sigma_j+\Delta_j$ represents the real measurements performed on the Bell state $\ket{\Phi^+_2}$.

To estimate the lower value of the witness when evaluated on 
a separable state $\rho_{AB}=\sum_{\lambda}p(\lambda)\langle a_j\rangle_\lambda \rho^B_{\lambda}$, 
accounting for the imperfections of self-testing, we are able to derive a steering inequality
\begin{align}
		W^{\rm noisy}_{\rm DI}=&\sum_{a,c,j} g_{c,j} a_j P(a, {\rm Yes}, c\,|x=j, \mathcal{B}, z=j)\nn\\
		=&\xinhui{\sum\limits_{\lambda, a, j}p(\lambda)\langle a_j\rangle_\lambda {\rm Tr}[\sum\limits_{c} g_{c,j} E_{\rm Yes}^{BB_0} \rho^B_\lambda \otimes\tau_{c,j}]}\nn\\
		=&\sum\limits_{\lambda, a, j}p(\lambda)\langle a_j\rangle_\lambda {\rm Tr}[\sum\limits_{c} E_{\rm Yes}^{BB_0} \rho^B_\lambda \otimes[\mathbb{I}+c (\tilde\sigma_j+\Delta_{j})]/2]\nn \\
		=&\sum\limits_{\lambda}p(\lambda)\sum\limits_{a, j} \langle a_j\rangle_\lambda {\rm Tr}[E_{\rm Yes}^{BB_0} \rho^B_\lambda \otimes (\tilde\sigma_j+\Delta_{j})]\nn \\
		=&W_{\rm DI}+\sum\limits_{\lambda}p(\lambda)\sum\limits_{a, j} \langle a_j\rangle_\lambda {\rm Tr} [E_{\rm Yes}^{BB_0} \rho^B_\lambda\otimes\Delta_{j}].
	\end{align}
Here $E^{BB_0}$ models the answer ``Yes" from Bob's arbitrary joint measurement $\mathcal{B}$, \xinhui{and $\tilde\sigma$ denotes the second term in Eq.~(S27)}. The third equality results from the relation $g_{c, j}=c=\pm 1$ \xinhui{ and $\tau_{c,j}=\frac{1}{2}(\mathbb{I}+c(\tilde\sigma_j+\Delta_j))$}. If self-testing is perfect, i.e., $f_j=\alpha_j=1$ and thus $\Delta_j=0$, then the above quantity recovers the ideal one $W_{\rm DI}$. 

Next, we analyse the noise range induced by imperfection of self-testing. \xinhui{Note first that
	\begin{align}
		&|\sum\limits_{\lambda}p(\lambda)\sum\limits_{a, j} \langle a_j\rangle_\lambda {\rm Tr} [E_{\rm Yes}^{BB_0} \rho^B_\lambda\otimes\Delta_{j}]| \nn \\
			\leq& \sum\limits_{\lambda}p(\lambda)\sum\limits_{j} |{\rm Tr} [E_{\rm Yes}^{BB_0} \rho^B_\lambda\otimes\Delta_{j}] |\nn \\
			\leq& \max_{\rho_\lambda}\sum\limits_{j} |{\rm Tr} [E_{\rm Yes}^{BB_0} \rho^B_\lambda\otimes\Delta_{j}]|.
			\end{align}
 It follows further from the positivity of the measurement element $E^{BB_0}_{\rm Yes}$ and states $\rho^B_\lambda$ that the partial trace $\tilde{\rho}^{B_0}_{\lambda}\equiv{\rm Tr_B}(E^{BB_0}_{\rm Yes} \rho_\lambda^B\otimes \mathbb{I})$ must be also a positive matrix. Thus, we are able to obtain 
 \begin{align}
 |{\rm Tr}[\Delta_j \tilde{\rho}^{B_0}_{\lambda}]| \leq |\lambda_j\cdot \lambda_{\max}(\tilde{\rho}^{B_0}_{\lambda})|=\sqrt{1-f_j}\lambda_{\max}(\tilde{\rho}^{B_0}_{\lambda})\leq \sqrt{1-f_j}.
 \end{align}
The first inequalities follows from the spectral decomposition of $\Delta_j$ with two eigenvalues $\lambda_{j}=\pm \sqrt{1-f_j}$ and there is a trivial bound $1$ for the quantity $\max_{\rho_\lambda}\lambda_{\max}(\tilde{\rho}^{B_0}_{\lambda})$ or $\max_{\rho_\lambda} \|Tr_B...\|$, as  the eigenvalues of all positive matrices $E^{BB_0}_{\rm Yes}$ and $\rho_\lambda$ are no larger than 1. In practice, if $E^{BB_0}_{\rm Yes}$ models the ``Yes'' from Bob's joint partial BSM $\mathcal{B}=\{\mathcal{B}_1,\mathbb{I}-\mathcal{B}_1\}$, where $\mathcal{B}_1=\ket{\Phi^+_2}\bra{\Phi^+_2}$ with $\ket{\Phi^+_2}=\frac{1}{\sqrt{2}}(\ket{00}+\ket{11})$, then there is $\max_{\rho_\lambda}\lambda_{\max}(\tilde{\rho}^{B_0}_{\lambda})=\max_{\rho_\lambda}\lambda_{\max}((\rho^{B}_{\lambda})^\top/2)\leq 1/2$. }

 Considering the worst case,  we obtain 
\begin{align}
 \xinhui{ W^{\rm noisy}_{\rm DI}\leq W_{\rm DI}-\max \sum_j \lambda_j \max_{\rho_\lambda}||{\rm Tr}_{B} (E_{\rm Yes}^{BB_0} \rho^B_\lambda\otimes \mathbb{I})||_1= W_{\rm DI}-\frac{1}{2}\sum_j(\sqrt{1-f_j})\leq 0.}
\end{align}
Thus, if $W^{\rm noisy}_{\rm DI}>0$ witnesses steerability conclusively, under the imperfection of self-testing.

For the class of Werner states given as 
\begin{align}
\rho=v\ket{\Psi^-_2}\bra{\Psi^-_2}+(1-v)\frac{\mathbb{I}}{4},
\end{align}
the implementation of Bob's partial BSM leads to

\begin{align}\label{noisySI}
W^{\rm noisy}_{\rm DI}(\rho)=\frac{1}{4}\left[3v-\sqrt{3}\right]-\frac{1}{2}
\sum\limits_{j=1,2,3}(\sqrt{1-f_j})\leq 0.
\end{align} 

\section{Experimental details}
In this section, we will give the details about the generation of the photon source, the construction of the partial BSM and the settings of the wave plates used in the self-testing stage.

\textit{Photon source-} In our experiment, the maximally entangled state $\ket{\Phi^+_2}=(\ket{00}+\ket{11})/\rt{2}$ is prepared through the SPDC process, where the pump laser has a repetition rate of 80 MHz, a central wavelength of 390 nm, and a pulse duration of 140 fs. A sandwich-like $\beta$-barium-borate crystal is configured in SPDC and a pair of the YVO4 crystal and LiNO3 crystal is used for temporal and spatial compensations~\cite{Zhang2015}.  To be specific, the computer basis $0, 1$ are encoded on the photon's horizontally polarized direction (H) and vertically polarized direction (V) respectively.

\begin{figure*}[htbp]
	\centering
	\includegraphics[width=\linewidth]{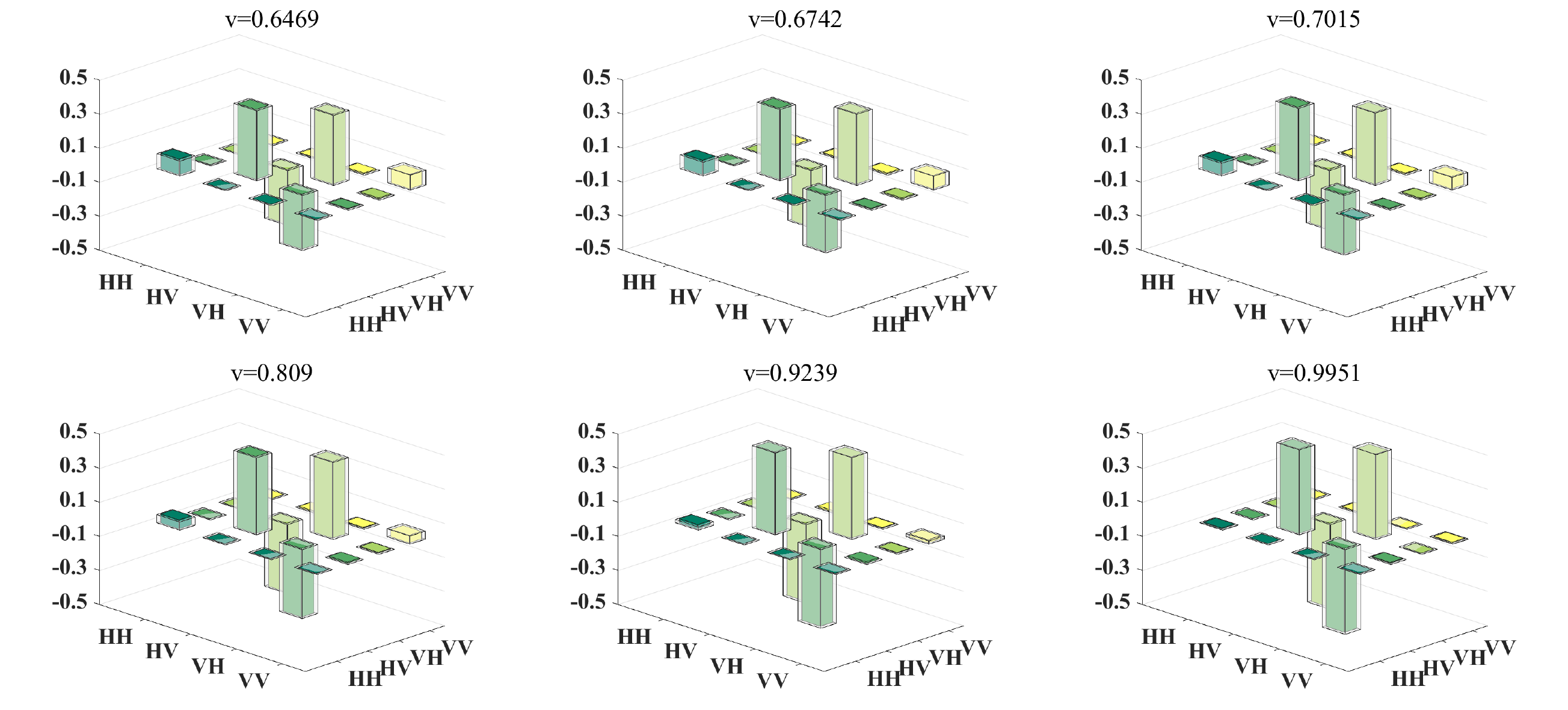}
	\caption{State tomography for Werner states. The real parts of the Werner states are shown as the colorful bars, and the correspondingly theoretical values are as the transparent bars. Each state is constructed from about $9,800,000$ photon pairs. }\label{fig:tomo}
\rule[-10pt]{13cm}{0.05em}
\end{figure*}

The singlet state $\ket{\Psi_2^-}=\frac{1}{\sqrt{2}}(\ket{01}-\ket{10})$ is prepared by re-encoding one photon's polarization $H$($V$) as $1$($0$) for state $\ket{\Phi_2^+}$ and slightly tilting the temporal compensation crystals YVO4 to add a phase $\pi$. In the experiment, we simulate the added white noise of the to-be-witnessed system $\rho_{AB}$ by flipping Alice's measurement, and the noise level $v$ is roughly estimated by the flipping probability $(1-v)/2$  \cite{Saunders2017}. By performing the standard quantum state tomography, we get the density matrix of the experimentally prepared state, which is approximated to the Werner state $\rho_W$ with visibility $v$. The real part of density matrices $\rho_{AB}$ and the proximate Werner states are shown in Fig.~\ref{fig:tomo}, and the corresponding fidelities are $0.9993(4)$, $0.9993(4)$, $0.9993(4)$, $0.9988(4)$, $0.9960(4)$ and $0.9959(1)$ respectively.

\textit{Partial BSM-} \xinhui{The measurement bases of BSM are in the form of  four Bell states $\{\ket{\Phi_2^\pm},\ket{\Psi_2^\pm}\}$, where $\ket{\Phi_2^\pm}=\frac{1}{\sqrt{2}}(\ket{00}\pm\ket{11}$) and $\ket{\Psi_2^\pm}=\frac{1}{\sqrt{2}}(\ket{01}\pm\ket{10})$. The partial BSM is implemented by $\mathcal{B}=\{\mathcal{B}_1,\mathbb{I}-\mathcal{B}_1\}$ where $\mathcal{B}_1=\ket{\Phi^+_2}\bra{\Phi^+_2}$. } In our experiment, we detect all two-photon coincidence of the eight APDs (D1~D8) in the BSM device, and category the results into four classes.
i) The coincidence happens between (1H, 2H) or (1V, 2V), the BSM resolves the \xinhui{$\ket{\Phi_2^+}$} state. ii) The coincidence happens between (1H, 2V) or (1V, 2H), the BSM resolves the \xinhui{$\ket{\Phi_2^-}$} state. iii) Both the two APDs in one output port fire, the BSM device detects the state \xinhui{$\ket{\Psi_2^+}$} or \xinhui{$\ket{\Psi_2^-}$}, and we can't tell two states apart. iv)The coincidence happens between (1H, 1V) or (2H, 2V) are attributed to the high-order emission noise or the imperfection of the HOM interference.

\begin{figure*}[htbp]
	\centering
	\includegraphics[width=0.9\linewidth]{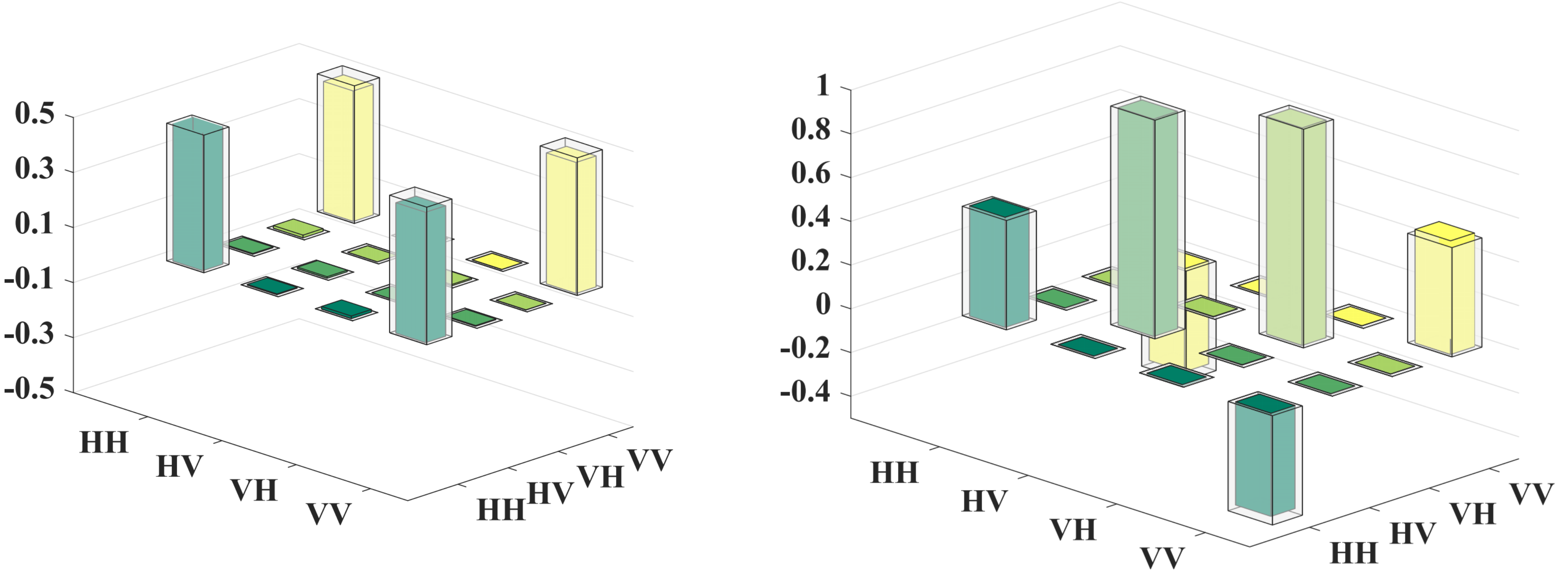}
	\caption{Measurement tomography for the partial BSM. The real part of the matrix $\widetilde{\mathcal{B}}_1^{ex}$ (the left histogram) and $\widetilde{\mathcal{B}}_2^{ex}$ (the right histogram) with the ideal theoretical values covered. }\label{fig:BSM}
	\rule[-10pt]{13cm}{0.05em}
\end{figure*}

A standard quantum measurement tomography is performed to estimate the detailed form of the experimentally implemented BSM. In the process, $36$ states, the tensor products of the eigenstates of the Pauli operators $\sigma_x$, $\sigma_y$ and $\sigma_z$ are prepared and sent to our partial BSM module. Then the maximum likelihood method is used to estimate the POVM elements. The fidelity between the experimentally constructed BSM and the ideal BSM is defined by the fidelity of quantum state: $F(\mathcal{B}^{ex},\mathcal{B})=(\sum_{j=1}^2w_i\sqrt{F_j})^2$, where $w_j=\frac{\sqrt{\tr{\mathcal{B}_j^{ex}}\tr{\mathcal{B}_j}}}{d}$, $F_j=F(\widetilde{\mathcal{B}}_j^{ex},\widetilde{\mathcal{B}}_j)$ is the fidelity between the normalized BSM elements $\widetilde{\mathcal{B}}_j^{ex}=\frac{\mathcal{B}_j^{ex}}{\tr{\mathcal{B}_j^{ex}}}$ and $\widetilde{\mathcal{B}}_j=\frac{\mathcal{B}_j}{\tr{\mathcal{B}_j}})$, and $\mathcal{B}_j^{ex}$ is the experimentally implemented BSM element. In our experiment, the overall fidelity of the partial BSM is $F=0.9831\pm0.0040$ and the purity of $\mathcal{B}_1$ is given by $P_1=Tr({\frac{\mathcal{B}_1^{ex}}{\tr{\mathcal{B}_1^{ex}}}}^2)=0.9547$. The estimated forms of the normalized POVM elements $\widetilde{\mathcal{B}}_j^{ex}$ are given in Fig.~\ref{fig:BSM}. The main errors are caused by the imperfection of the HOM-type interference, where the photons coming from different sources are not completely indistinguishable.

\begin{table}
	\centering
	\begin{tabular}{|c|c|c|c|c|c|} \hline
		
		\multicolumn{3}{|c|}{Bob}& \multicolumn{3}{|c|}{Charlie}\\
		\hline
		observable & QWP($^\circ$) & HWP($^\circ$) & observable & QWP($^\circ$) & HWP($^\circ$) \\ \hline
		X+Z & 22.5 & 11.25 & X & 45 & 22.5 \\ \hline
		X-Z & -22.5 & -56.25 & Z & 0 & 0\\ \hline
		X+Y & 45.00 & 33.75 & X & 45.00 & 22.50 \\ \hline
		X-Y & 45.00 & 11.25 & Y & 0 & 22.5 \\ \hline
		Y+Z & 0 & 11.45 & Y &0 & 22.5 \\ \hline
		Y-Z & 0 & -56.25 & Z &0 & 0 \\ \hline
	\end{tabular}
	\caption{Detailed parameters of wave plates set for Charlie and Bob to do self-testing. The $X$, $Y$ and $Z$ denote the Pauli operators $\sigma_x$, $\sigma_y$ and $\sigma_z$, respectively. }
	\label{measurement_setting}
\end{table}

\section{In comparison  to DI verification of entanglement}

\begin{figure}[ht]
	\centering
	\includegraphics[scale=0.4]{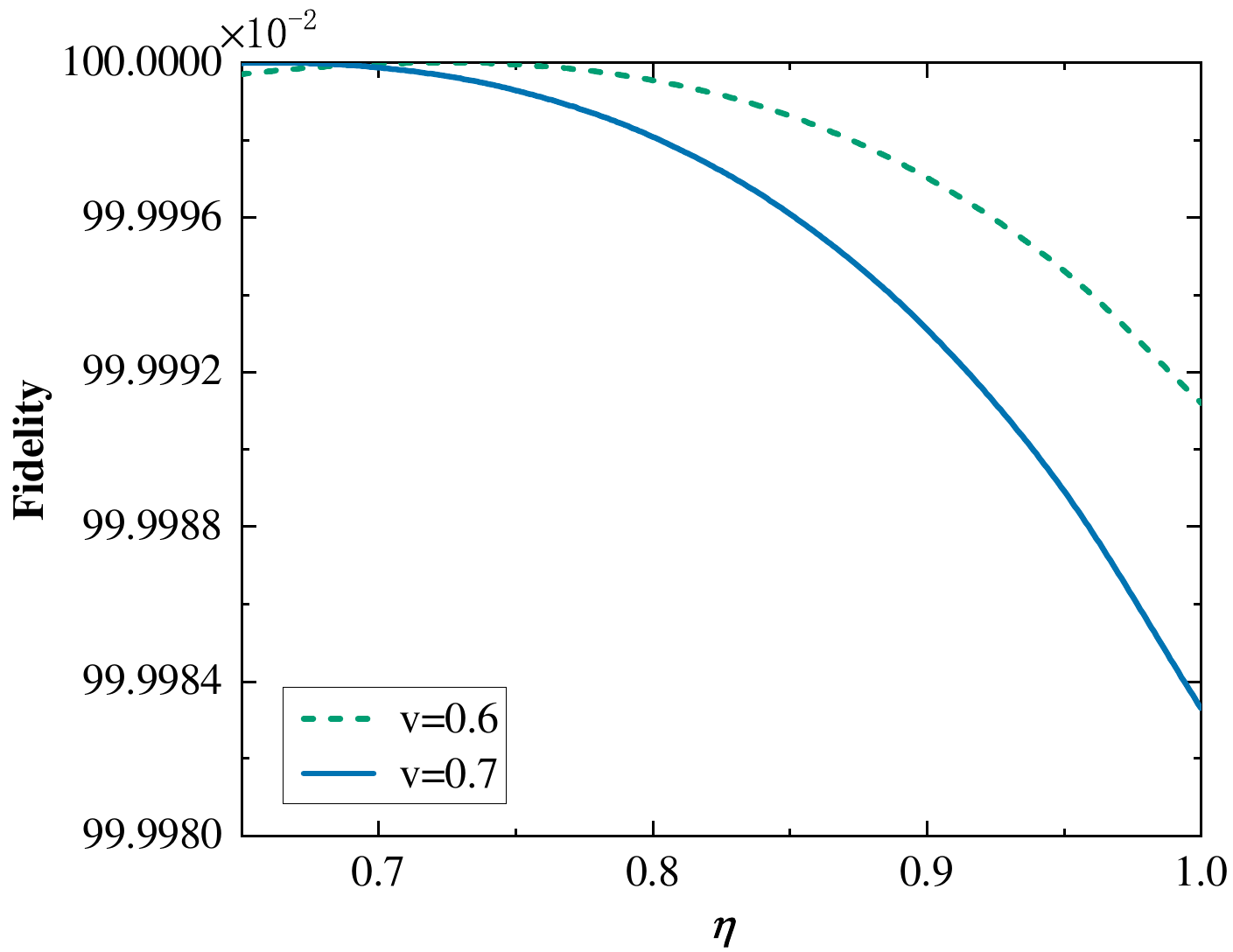}
	\caption{The average fidelity of self-testing for DI verification of entanglement of Werner states with $v=0. 6$ and $0.7$ derived in~\cite{Bowles2018b}. The fidelity of Pauli observables requires near-perfect self-testing to faithfully complete DI verification task, which is hard to reach within current technology.}\label{fig:PRA}
\end{figure}

By contrast, it was discussed in~\cite{Bowles2018b} that to faithfully verify entanglement for Werner states device-independently. One can certify entanglement if  
\begin{align}
\mathcal{I} =&\frac{1}{16}\big((1-3v)\eta^2+2\eta(1-\eta)+\frac{1}{4}(1-\eta)^2\big) \nn\\
\leq &\xinhui{-12[\big(\sqrt{2(1-\widehat{f})}+1-\widehat{f}\big)^2+\sqrt{2(1-\widehat{f})}+(1-\widehat{f})]},
\end{align}
where $\eta$ is the \xinhui{visibility} of preparing the Bell state $\ket{\Phi^+_2}$ being self-tested, and we use \xinhui{average fidelity $\widehat{f}=1-\theta^2$} to replace the original one obtained in~\cite{Bowles2018b}. The fidelity required to verify entanglement for different values of $\eta$ with $v=0.6$ and $0.7$ is plotted in Fig.~\ref{fig:PRA}. It is obvious that even for $\eta=1$,  it requires extremely high fidelity, i.e.$\widehat{f}>0.99999$ for $v=0.6$ and $\widehat{f}>0.99998$ for $v=0.7$ which are hard to realize in experiments, while our result derived in Eq.~(\ref{noisySI}) allows the fidelity of around 0.997, which is a significant reduction and attainable in current experiments.

\end{document}